\documentclass[twocolumn,showpacs,amsmath,amssymb,pra,floatfix]{revtex4-1}

\usepackage{graphicx}
\usepackage{dcolumn}
\usepackage{bm}
\usepackage{bbm}
\usepackage{multirow,amssymb,amsbsy,amsmath}
\usepackage{stmaryrd}

\newcommand{\ddim}{\udelta\kern0.1em}

\newcommand{\beikonst}[2]{\left( #1 \right)_{\kern-0.2em #2}}
\newcommand{\tr}[2][]{\text{Tr}_{#1}\left\{#2\right\}}

\newcommand*{\ket}[1]{\mathopen{|}#1\mathclose{\rangle}}

\newcommand{\comu}[2]{\left[#1,#2\right]}

\newcommand{\ketbra}[1]{\mathopen{|}#1\mathclose{\rangle}\hspace{-0.25em}\mathopen{\langle}#1\mathclose{|}}
\newcommand{\ketbrap}[2]{\mathopen{|}#1\mathclose{\rangle}\hspace{-0.25em}\mathopen{\langle}#2\mathclose{|}}

\begin{document}


%
%

\title{Tailored jump operators for purely dissipative quantum magnetism}

\author{Hendrik Weimer}%
\email{hweimer@itp.uni-hannover.de}
\affiliation{Institut f\"ur Theoretische Physik, Leibniz Universit\"at Hannover, Appelstr. 2, 30167 Hannover, Germany}

\date{\today}%

\begin{abstract}
  
  I propose an archtitecture for the realization of dissipative
  quantum many-body spin models. The dissipative processes are
  mediated by interactions with auxiliary particles and lead
  to a widely tunable class of correlated quantum jump
  operators. These findings enable the investigation of purely
  dissipative spin models, where coherent dynamics is entirely absent.
  I provide a detailed review of a recently introduced variational
  method to analyze such dissipative quantum many-body systems, and I
  discuss a specific example in terms of a purely dissipative
  Heisenberg model, for which I find an additional disordered phase
  that is not present in the corresponding ground state phase diagram.

\end{abstract}


\maketitle

\section{Introduction}

Emerging quantum technologies such as quantum simulation or quantum
metrology inherently require control over nonclassical correlations
such as quantum entanglement. For a long time, a central belief has
been that in order to realize such technologies, it would be
absolutely necessary to perfectly isolate the quantum system of
interest from its environment, as otherwise there would be unavoidable
decoherence destroying the nonclassical states. However, as has been
demonstrated in a series of landmarking theoretical and experimental
works
\cite{Diehl2008,Verstraete2009,Weimer2010,Barreiro2011,Krauter2011,Lin2013,Shankar2013},
it is possible to use carefully engineered couplings to the
environment to dissipatively prepare interesting quantum many-body
states as the stationary states of the evolution of such an open
quantum system.

These initial results have sparked a huge interest in the study of
dissipative quantum many-body systems. Besides investigation of the
aforementioned dissipative quantum state engineering
\cite{Diehl2010,Alharbi2010,Watanabe2012,Ates2012,Rao2013,Carr2013a,Lemeshko2013a,Weimer2013a,Hofer2013,Otterbach2014},
there is also a large body of works investigating their
fundamental properties. In particular, identifying and understanding
phase transitions in dissipative quantum systems has emerged as a
central topic in this context \cite{Hartmann2010,Baumann2010,Nagy2010,Diehl2010a,Tomadin2011,Lee2011,Kessler2012,Honing2012,Honing2013,Horstmann2013,Torre2013,Sieberer2013,Qian2013,Lee2013,Carr2013,Joshi2013,Malossi2014,Marcuzzi2014,Lang2015,Marino2016,Marcuzzi2016}.

The theoretical analysis of dissipative quantum many-body systems is
significantly more challenging than for equilibrium systems due to the
lack of an underlying concept corresponding to the partition
function. Consequently, the available tools have been quite
limited. For one-dimensional problems, methods related to the density
matrix renormalization group have been successfully used
\cite{Honing2012,Honing2013,Pizorn2013,Transchel2014,Cui2015,Mascarenhas2015,Mendoza-Arenas2016,Werner2016},
however their extension towards higher dimensional problems remains a
outstanding challenge. In higher dimensions, most analyses have been
limited to mean-field treatments
\cite{Tomadin2010,Diehl2010a,Tomadin2011,Liu2011,Lee2011,Lee2012a,Qian2013,Lee2013,Jin2013,Poletti2013,Vidanovic2014},
however, recent results cast severe doubts on the validity of such a
mean-field decoupling
\cite{Hoening2014,Weimer2015a,Mendoza-Arenas2016,Overbeck2016}, even
in large spatial dimensions. Generic non-equilibrium methods such as
the Keldysh formalism can be extended to open systems
\cite{Sieberer2013,Maghrebi2016}, but are notoriously difficult to
treat in the presence of strong interactions \cite{Eckstein2009}. As a
result, there is increasing activity to find novel computational
approaches that do not suffer from these limitations
\cite{Finazzi2015,Li2016,Jin2016}. One promising route is a
variational principle for open quantum systems \cite{Weimer2015},
which has given quantitatively reliable results for a variety of
different systems
\cite{Weimer2015a,Overbeck2016,Kaczmarczyk2016,Lammers2016}, and which
appears to correctly describe phase transitions above the upper
critical dimension \cite{Overbeck2016a}.

The physical realizations that are being discussed in the context of
dissipative quantum many-body systems are quite diverse. On the one
hand, a wide range of works investigates the coupling of atoms or ions
to metastable electronic excitations
\cite{Baumann2010,Carr2013,Malossi2014,Weber2015}. On the other hand,
there is also a very strong activity in the context of solid state
systems, e.g., in the context of exciton-polariton condensates
\cite{Kasprzak2006,Lai2007,Amo2009}, solid state cavity arrays
\cite{Tomadin2010,Liew2013,LeBoite2013,Aron2016,Kimchi-Schwartz2016},
or nitrogen-vacancy defect centers in diamond
\cite{Robledo2011,Bar-Gill2012,Weimer2012,Weimer2013}. A key point in
all these investigations is that the steady state of the open quantum
system is not a thermal state with respect to the system
Hamiltonian. One common way to realize this is to consider an explicit
time-dependent driving field, e.g., an external laser, which is why
such systems are sometimes refered to as being
``driven-dissipative''. However, it is important to note that
non-Markovian systems offer similar possiblities
\cite{deVega2016,Iles-Smith2016}, and hence it is possible to realize
interesting dissipative quantum many-body models also without the help
of external driving fields.

Nevertheless, it remains a central challenge to realize dissipative
processes in a controlled way. In the context of driven-dissipative
settings, the dissipation usually arises from coupling the eigenstates
of the system to the vacuum of the electromagnetic field, hence the
dissipation naturally acts in the eigenbasis of the system. In many
cases, however, it would be desirable to engineer other dissipation
channels, e.g., for the production of steady state coherence between
the eigenstates. While there are proposals to realized largely
arbitrary dissipation channels in the framework of universal quantum
simulation architectures
\cite{Weimer2010,Weimer2011,Weimer2013a,Georgescu2014}, it is highly
desirable to implement tunable dissipation without requiring such a
detailed level of control over the setup.

In this paper, I disscuss an architecture for the realization of
largely arbitrary dissipation channels. The central strategy is to
turn coherent interactions with auxiliary particles into effective
dissipative elements for the reduced system under consideration. I
will discuss the implementation of single-particle and two-particle
operators describing dissipative quantum jumps, focussing on the case
where the resulting effective dynamics is purely dissipative and
coherent dynamics is entirely absent. Crucially, such purely
dissipative systems are still fully quantum if the associated quantum
jump operators act in different eigenbases \cite{Lang2015}. As a
specific example, I will focus on dissipative quantum magnetism in the
form of a dissipative Heisenberg model. I will analyze the phase
diagram of this model using the variational principle for open quantum
systems and comment on the differences to the ground state phase
diagram.

\section{Tailoring quantum jump operators}

\begin{figure}[t]
  \includegraphics{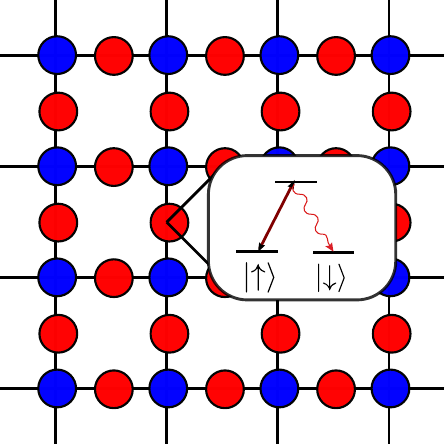}
  \caption{Setup of the system where atoms of the spin ensemble (blue)
    are interspersed with auxiliary atoms (red) needed to mediate
    dissipative couplings. The auxiliary atoms have one spin state
    decaying into the other by means of optical pumping via an
    electronically excited state.}
  \label{fig:setup}
\end{figure}

Here, we are interested in realizing dissipative spin models for an
ensemble of atoms or molecules loaded into an optical lattice with
uniform filling.  We will be investigating a case where the
dissipative elements of the dynamics are mediated by a set of
auxiliary atoms, see Fig.~\ref{fig:setup}. The starting point of the
analysis is the microscopic quantum master equation in Lindblad form,
which includes both the spin ensemble and the auxiliary atoms, given
by
\begin{align}
  \frac{d}{dt}\rho=-i[H,\rho]+ \sum\limits_{i}\mathcal{D}(c_i),
 \label{eq:mastereq}
\end{align}
where $\rho$ is the density operator describing the state of the
system, $H$ is the Hamiltonian responsible for the coherent part of
the dynamics, and the dissipative terms are given in terms of quantum
jump operators $c_i$ by
\begin{equation}
  \mathcal{D}(c_i) = c_i\rho
    c_i^{\dagger}-\frac{1}{2}\left(c_i^\dagger c_i\rho + \rho c_i^\dagger c_i\right).
\end{equation}
In the following, we will obtain a reduced description of the dynamics
only for the atoms in the spin ensemble.

\subsection{Effective operator formalism}

The general setup we are considering has dissipative terms only within
the auxiliary atoms. Therefore, we aim for an effective description
where the dynamics of the auxiliary atoms is integrated out. To be
specific, we assume that the auxiliary atoms are two-level systems
with one state decaying into the other; a straightforward realization
is optical pumping by laser coupling one of the two levels to a fast
decaying electronic excitation, see Fig.~\ref{fig:setup}. Within this
assumption, we will employ the effective operator formalism
\cite{Reiter2012} to derive an effective quantum master equation for
the spin ensemble.

Within the effective operator formalism, the resulting master equation
is given by
\begin{equation}
  \dot\rho = -i\comu{H^\text{eff}}{\rho} + \sum\limits_k \mathcal{D}(c^\text{eff}_k),
\end{equation}
with the effective Hamiltonian $H_\text{eff}$ and effective jump
operators $c^k_\text{eff}$ being given by
\begin{align}
  H_\text{eff} &= H_g -\frac{1}{2}\sum\limits_k V^-_k\left[\tilde{H}_k^{-1} + \left(\tilde{H}_k^{-1}\right)^\dagger\right]V^+_k\\
  c^\text{eff}_k &= c_k\tilde{H}_k^{-1}V_k^+.
\end{align}
Here, $H_g$ is the Hamiltonian within the ground state manifold, i.e.,
involving only states that are not subject to dissipation. $V_k^+$ and
$V_k^-$ are interaction terms that couple an ensemble atom to the
decaying state or away from it, respectively. Finally, $\tilde{H}_k$
is the non-Hermitian Hamiltonian for the $k$th auxiliary atom, given by
\begin{equation}
  \tilde{H}_k = H_k^e -\frac{i}{2}c_k^\dagger c_k,
\end{equation}
which $H_k^e$ being the part of the Hamiltonian that couples only
within the manifold of states subject to dissipation. Note that
although $\tilde{H}_k$ is non-Hermitian, the overall effective
Hamiltonian $H^\text{eff}$ is always Hermitian.

In the following, we will be interested the case where the effective
dynamics is purely dissipative. To this end, we consider situations
where $H_g = 0$, which can be realized by ensuring that there are no
interactions within the spin ensemble, while possible effective terms
corresponding to local operators can be canceled using suitable local
potentials. The contribution to the effective Hamiltonian of the
auxiliary atoms vanishes in case the non-Hermitian Hamiltonian is
purely imaginary; this occurs when the two states of the auxiliary
atom have the same energy in a suitable rotating frame.

\subsection{Single-particle jump operators}

As a first step, let us discuss the requirements for the realization
of arbitrary single-particle jump operators. To be specific, we will
exemplify the proposed architecture by considering the realization of
the jump operator $\ketbrap{-}{+}$, as this jump operator is not
acting within the eigenbasis of the spin ensemble and therefore
represents are particularly challenging case.

Without loss of generality, we can assume that the spin up state of
the auxiliary atom is decaying into the spin down state, hence the
microscopic jump operator is given by $c_k = \sqrt{\gamma}
\sigma_k^-=\sqrt{\gamma}[\sigma_k^x-i\sigma_k^y]/2$ in terms of the
Pauli matrices $\sigma_k^\alpha$, with $\gamma$ being the associated
decay rate. Then, we will be insterested in the case where the
interaction Hamiltonian between the $i$th atom in the spin ensemble
and the $k$th auxiliary atoms is given by
\begin{equation}
  H_{ik} = E_0\ketbrap{-}{+}_i \sigma^+_k + \text{H.c.} = \frac{E_0}{2}\left[\sigma_i^z\sigma_k^x+\sigma_i^y\sigma_k^y\right],
\end{equation}
with $E_0$ being an energy scale related to the concrete physical
implementation of the interaction. Such tunable spin interactions can
be efficiently realized in a wide range of physical systems, most
notably Rydberg-dressed atoms \cite{Glaetzle2015,vanBijnen2015} and
rotational excitations of ultracold polar molecules
\cite{Micheli2006,Gorshkov2011,Gorshkov2013}. Within the context of
the effective operator formalism, we find that the term containing
$\sigma^+_k$ corresponds to $V_k^+$, while its Hermitian conjugate is
identical to $V_k^-$. Then, we find for the effective jump operator
\begin{equation}
  c^\text{eff}_i = \frac{E_0}{\sqrt{\gamma}} \ketbrap{-}{+}_i \ketbra{\downarrow}_k.
\end{equation}
In leading order in $E_0/\gamma$, the auxiliary atoms are confined to
the spin down state and hence the auxiliary atom can be factorized out
from the dynamics, leaving only the desired jump operator
$\ketbrap{-}{+}$.

\subsection{Multi-particle jump operators}

\label{sec:multi}

Now, let us turn to the realization of two-body jump operators. In the
same way as the realization of single-particle jump operators requires
two-body interactions including one of the auxiliary atoms, two-body
jump operators require the presence of three-body interactions. Such
three-body terms can arise in a variety of different contexts, e.g.:
(i) Strong blockade effects in Rydberg-dressed atoms lead to
higher-order interaction terms \cite{Honer2010}. (ii) Canceling the
two-body interaction leave three-body interactions as the leading
interaction term \cite{Buchler2007a}. (iii) Time-dependent driving of
a system can lead to three-body interactions in the rotating frame of
the driving \cite{Lee2016}. In the following, we will not make detailed
assumption about the underlying physical mechanism to realize the
three-body interaction terms, allowing to retain a general treatment
of the emerging dissipative many-body dynamics.

As an example, let us study the implementation of a correlated jump
operator of the form $\ketbrap{\psi_+}{\psi_-}$, where the states
$\ket{\psi_\pm}$ are the Bell states
\begin{equation}
  \ket{\psi_\pm} = \frac{1}{\sqrt{2}}(\ket{\uparrow\downarrow}\pm\ket{\downarrow\uparrow}.
  \label{eq:psi}
\end{equation}
As this jump operator pumps the system towards the maximally entangled
state $\ket{\psi_+}$, it can be seen as a building block for the
efficient generation of entangled quantum states using dissipation. In
close analogy to the preceding subsection, we can realize such a jump
operator with the help of the auxiliary atoms. Specifically, we find
this jump operator being realized when the interaction Hamiltonian
$H_{ijk}$ between two atoms of the spin ensemble $i$ and $j$ and one
auxiliary atom $k$ takes the form
\begin{align}
  H_{ijk} &=  E_0\ketbrap{\psi_+}{\psi_-} \sigma^+_k + \text{H.c.}\\\nonumber &= \frac{E_0}{4}\left[\sigma_i^x\sigma_j^x\sigma_k^x+\sigma_i^y\sigma_j^y\sigma_k^x-\sigma_i^z\sigma_j^z\sigma_k^x + \sigma_k^x\right].
\end{align}
Calculation of the effective jump operators then leads to the desired
result.

\section{Variational analysis of purely dissipative Heisenberg models}

We will now turn to the analysis of purely dissipative spin models
that can be realized using the techniques laid out in the previous
section. However, before turning to a detailed analysis of the models,
let us go through a detailed introduction to the variational principle
presented in \cite{Weimer2015}, which will serve as the basis for our
investigation.

\subsection{Introduction to the variational principle for dissipative quantum many-body systems}

In general, finding the exact solution to the steady state of the
quantum master equation is an exponentially hard problem. Therefore,
we will aim for an approximate solution that can be calculated
efficiently. To this end, we perform a variational paramterization of
the density matrix. As the first step, we take a variational basis in
terms of tensor products of single atom density matrices. As a further
simplification, we assume that all single atom density matrices are
identical.  Then, we have
  \begin{align}
    \label{eq:product}
    \rho &= \rho_0 \otimes \rho_0 \otimes \cdots  \\
    \rho_0 &= \frac{1}{2}\left(1+\alpha_x\sigma_x + \alpha_y \sigma_y + \alpha_z\sigma_z\right).
  \end{align}
  Here, $\alpha_x$, $\alpha_y$, and $\alpha_z$ are our (real)
  variational parameters, which we will vary to find the best possible
  approximation to the steady state. Consequently, we have boiled down
  the full quantum many-body problem down to the computation of three
  variational parameters. Note that we have already incorporated the
  requirement of the trace being identical to one into the
  construction of $\rho_0$, as the Pauli matrices are
  traceless. However, the constraint that the density matrix must be
  non-negative leads to the condition
  $\alpha_x^2+\alpha_y^2+\alpha_z^2\leq 1$, which has to be accounted
  for within the numerical optimization.

  As the underlying variational principle, we want to minimize a
  suitable norm $||d/dt \rho||$, as the norm going to zero indicates
  that we have found the true steady state. Here, the suitable norm is
  given by the trace norm, i.e., $\tr{|d/dt \rho|}$
  \cite{Weimer2015}. However, computing the variational norm is still
  an exponentially hard task, even for our rather simple ansatz
  discussed above. Therefore, we minimize a slightly different
  variational functional that provides an upper bound. In the case of
  our ansatz, this upper bound can be given as
  \begin{equation}
    \label{eq:upper}
    ||\frac{d}{dt} \rho|| \leq \sum\limits_{\langle ij\rangle} ||\frac{d}{dt}\rho^{(ij)}||,
  \end{equation}
  where $d/dt \rho^{(ij)}$ is the time derivative of the reduced system
  consisting of only two atoms \cite{Weimer2015}. For system sizes
  where a direct comparison can be made, taking the upper bound
  instead of the full problem leads to very similar results
  \cite{Weimer2015a}.  Crucially, the reduced time derivative is given
  by a $4\times 4$ matrix, which can be easily diagonalized. In the
  case of a homogeneous system, all terms of the sum are identical,
  meaning that we have to minimize the norm of a single $4\times 4$
  matrix.

  In the most general case, the contribution to the operator $d/dt
  \rho^{(ij)}$ consists of three different parts,
  \begin{equation}
    \frac{d}{dt}\rho^{(ij)} = \dot\rho^{(ij)}_{loc} + \dot\rho^{(ij)}_{int} + \dot\rho^{(ij)}_{MF}.
    \label{eq:dotrhoparts}
  \end{equation}
  The first term, $\dot\rho^{(ij)}_{loc}$, refers to terms that act
  only within the local Hilbert space of a single site. The second
  term, $\dot\rho^{(ij)}_{int}$ refers to two-particle operators
  describing coherent interactions and correlated jump operators that
  act only on particles $i$ and $j$. These first two terms are
  identical to the ones found in the microscopic quantum master
  equation. The third term, $\dot\rho^{(ij)}_{MF}$ is slightly more
  subtle and corresponds to a mean-field treatment of the sites
  surrounding $i$ and $j$, given by
  \begin{equation}
    \dot\rho^{(ij)}_{MF} = \sum\limits_k \tr[k]{\comu{H_{ik} + H_{jk}}{\rho^{(ijk)}} + \mathcal{D}(c_{ik}) + \mathcal{D}(c_{jk})},
    \label{eq:dotrhomf}
  \end{equation}
  where $\rho^{(ijk)} = \rho^{(i)} \otimes \rho^{(j) }\otimes
  \rho^{(k)}$ is the reduced density operator for three sites. If we
  compose the interaction term $H_{ik}$ into operators of the form
  $H_{ik} = A_i \otimes B_k$, we can write their contribution to $
  \dot\rho^{(ij)}_{MF}$ as
  \begin{equation}
    \tr[k]{\comu{H_{ik}}{\rho^{(ijk)}}} = \tr{B_k\rho_k}\comu{A_i \otimes 1_j}{\rho^{(ij)}},
  \end{equation}
  which describes a mean-field decoupling of the interaction
  \cite{Weimer2008b}. Similar expressions can also be found in the case
  of correlated jump operators acting on multiple sites at the same
  time.

  The sum over $k$ in Eq.~(\ref{eq:dotrhomf}) runs over all sites
  adjacent to $i$ and $j$ and therefore contains $2(z-1)$ terms, where
  $z$ is the coordination number of the underlying lattice. Crucially,
  this results in the mean-field term being dominant in the limit of
  large $z$. If we want to capture, for instance, the changes in phase
  diagrams with changing $z$, it is therefore convenient to
  renormalize all coupling constants $\lambda$ corresponding to
  two-particle operators (both coherent and dissipative) according to
  $\lambda = \lambda'/(z-1)$. Then, the mean-field term remains
  constant in the large $z$ limit. However, this does not apply to the
  second term $\dot\rho^{(ij)}_{int}$, which decays like $1/(z-1)$ in
  this limit. As this term captures the buildup of correlations, this
  means that correlations become irrelevant in the limit of large $z$
  and the product state solution becomes exact.

  The variational approach has proven to be especially successful for
  the analysis of dissipative phase transitions
  \cite{Overbeck2016a}. The unique advantage is the possibility to
  interpret the variational norm in close correspondence to the free
  energy functional for equilbrium systems. Then, in close analogy to
  Landau theory for equilibrium transitions, one can formally perform
  a series expansion of the variational norm in the order parameter
  $\phi$ of the dissipative phase transition. A common situation is
  that due to symmetry reasons, the odd powers of the expansion vanish
  and one is left with a $\phi^4$ theory according to
  \begin{equation}
    ||\dot\rho^{(ij)}|| = u_0 + u_2 \phi^2 + u_4 \phi^4 + O(\phi^6).
  \end{equation}
  Such a $\phi^4$ theory has a phase transition at a critical value of
  $u_2=0$, going from a disordered phase with $\phi = 0$ for $u_2 > 0$
  to an ordered phase with $\phi \ne 0$ for $u_2 < 0$. Close to the
  phase transition, the order parameter behaves according to $\phi
  \sim u_2^{1/2}$, from which we can identify the critical exponent
  $\beta=1/2$.

\subsection{Application to purely dissipative Heisenberg models}

The Heisenberg model is a paradigmatic model for the study of quantum
magnetism. Here, we will be interested in a particular variant of a
spin $1/2$ model, where the coupling constant along two axes is
identical, which is also known as the XXZ model. Then, the Hamiltonian
of the equilibrium model is given by
\begin{equation}
  H = -J\sum\limits_{\langle ij\rangle} \sigma_x^{(i)} \sigma_x^{(j)} + \sigma_y^{(i)} \sigma_y^{(j)} + (1-\lambda)\,\sigma_z^{(i)} \sigma_z^{(j)}.
\end{equation}
Constructing a dissipative analog of this model can be done in
different ways, but in general, the dissipative model should exhibit
the same symmetries as the equilibrium one. This means that for
$\lambda = 0$, the model exibits an $SU(2)$ symmetry corresponding to
global rotation around any axis, while for $\lambda \ne 0$ this
symmetry is partially lifted and only a $U(1)$ symmetry corresponding
to rotations around the $z$ axis remains. Here, we have chosen the
interaction in the $XY$ plane to be ferromagnetic, however, an
underlying mirror symmetry along the $z$ axis results in the
antiferromagnetic case having the same spectrum \cite{Yang1966}.

The way we construct the dissipative Heisenberg model is to consider
two sets of jump operators. The first set of jump operators pumps the
system into a dark state where the $SU(2)$ symmetry is spontaneously
broken. In the ferromagnetic case, these jump operators have the form
\begin{align}
  c_{ij} &= \ketbrap{\uparrow\uparrow}{\psi_-}_{ij} =
\frac{1}{\sqrt{2}}  \left(\begin{array}{cccc}
    0 & 1 & -1 & 0\\
    0 & 0 & 0 & 0\\
    0 & 0 & 0 & 0\\
    0 & 0 & 0 & 0\end{array}\right)\\
  c_{ij}'&= \ketbrap{\uparrow\uparrow}{\psi_-}_{ij}= \frac{1}{\sqrt{2}} \left(\begin{array}{cccc}
    0 & 0 & 0 & 0\\
    0 & 0 & 0 & 0\\
    0 & 0 & 0 & 0\\
    0 & 1 & -1 & 0\end{array}\right)\\
  c_{ij}'' &= \ketbrap{\psi_+}{\psi_-}_{ij},=  \frac{1}{2} \left(\begin{array}{cccc}
    0 & 0 & 0 & 0\\
    0 & 1 & -1 & 0\\
    0 & 1 & -1& 0\\
    0 & 0 & 0 & 0\end{array}\right)
\end{align}
with $\ket{\psi_\pm}$ defined as in Eq.~(\ref{eq:psi}). This choice of
jump operators ensures that the quantum master equation is
$SU(2)$-symmetric, equally pumping the system into the three possible
ground states of the ferromagnetic Heisenberg interaction. It is
straightforward to check that the dark states of this set of jump
operators are ferromagnetic states with all spins pointing in the same
direction, but the direction itself can lie anywhere on the Bloch
sphere. Finally, it is worth mentioning that it is also possible to
construct the appropriate jump operators to implement the
antiferromagnetic model. However, in the antiferromagnetic case the
fluctuations created by the dissipation are so strong that they always
drive the system towards an unpolarized state.

The second set of jump operators will now lift the $SU(2)$ symmetry in
the same way the symmetry is lifted in the eqilibrium model. Here, we
consider the set of jump operators given by
\begin{align}
  c_{ij} &= \sqrt{\lambda}\ketbrap{\uparrow\downarrow}{\uparrow\uparrow}_{ij}= \sqrt{\lambda}  \left(\begin{array}{cccc}
    0 & 0 & 0 & 0\\
    1 & 0 & 0 & 0\\
    0 & 0 & 0 & 0\\
    0 & 0 & 0 & 0\end{array}\right)\\
  c_{ij}' &= \sqrt{\lambda}\ketbrap{\downarrow\uparrow}{\uparrow\uparrow}_{ij}=  \sqrt{\lambda}\left(\begin{array}{cccc}
    0 & 0 & 0 & 0\\
    0 & 0 & 0 & 0\\
    1 & 0 & 0 & 0\\
    0 & 0 & 0 & 0\end{array}\right)
\end{align}
\begin{align}
  c_{ij}'' &= \sqrt{\lambda}\ketbrap{\uparrow\downarrow}{\downarrow\downarrow}_{ij}= \sqrt{\lambda} \left(\begin{array}{cccc}
    0 & 0 & 0 & 0\\
    0 & 0 & 0 & 1\\
    0 & 0 & 0 & 0\\
    0 & 0 & 0 & 0\end{array}\right)\\
  c_{ij}''' &= \sqrt{\lambda}\ketbrap{\downarrow\uparrow}{\downarrow\downarrow}_{ij}= \sqrt{\lambda} \left(\begin{array}{cccc}
    0 & 0 & 0 & 0\\
    0 & 0 & 0 & 0\\
    0 & 0 & 0 & 1\\
    0 & 0 & 0 & 0\end{array}\right).
\end{align}
Crucially, these jump operators have two antiferromagnetic dark
states, $\ket{\uparrow\downarrow}$ and $\ket{\downarrow\uparrow}$. It
is straightforward to decompose all jump operators into the required
elementary interaction terms as discussed in Sec.~\ref{sec:multi}.

We will now perform the variational analysis of the dissipative Heisenberg
model. Most importanly, there is no Hamiltonian in the problem, hence
all the quantum features of the model arise from the non-commuting
quantum jump operators. Additionally, all jump operators involve
two-site operators, i.e., the $\dot\rho^{(ij)}_{loc}$ term in
Eq.~(\ref{eq:dotrhoparts}) is zero.

We will perform the variational analysis in two parts. First, we
investigate the behavior for small values of $\lambda$, where we
expect the existence of an XY phase, in which the $U(1)$ symmetry is
spontaneously broken. Second, we look at the limit of large $\lambda$,
where the model is expected to realize an Ising antiferromagnet with
broken $Z_2$ symmetry. To be explicit, we always focus on
three-dimensional models on a cubic lattice, as in higher dimensions
the product state ansatz of Eq.~(\ref{eq:product}) becomes more
reliable \cite{Overbeck2016a}. The minimization of the variational
norm is done numerically using standard nonlinear optimization
techniques.

\begin{figure}[t]
  \includegraphics[width=\linewidth]{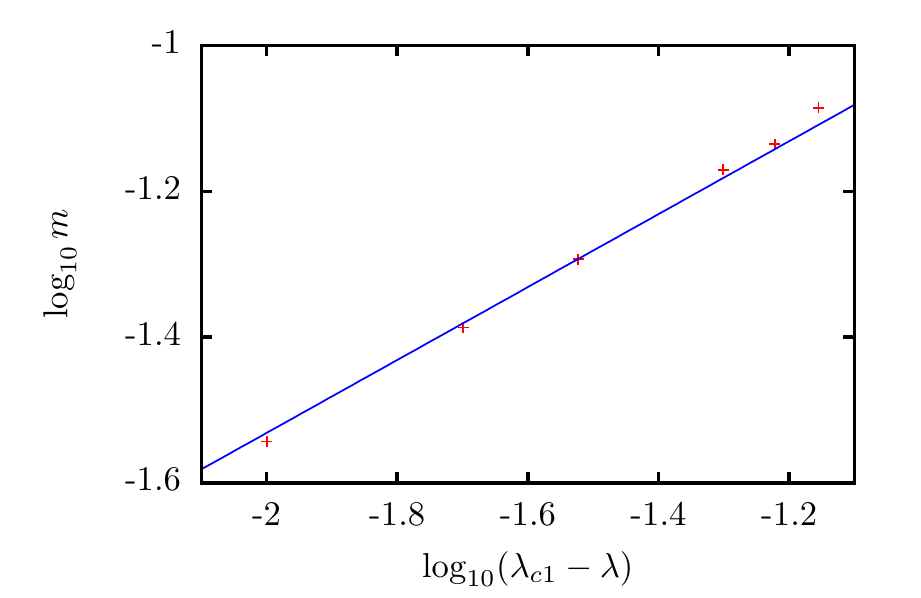}
  \caption{Critical behavior of the dissipative phase transition close to $\lambda_{c1}=1/2$
    between the XY and disordered phase. The fit represents a critical
    scaling of the order parameter according to $m \sim
    (\lambda_{c1}-\lambda)^{1/2}$.}
\label{fig:phase1}
\end{figure}

As noted before, for $\lambda=0$ any ferromagnetic state with all
spins pointing in the same direction is an exact dark state of the
quantum master equation. For any finite value of $\lambda>0$, however,
we find that the variational minimum is realized for $\langle
\sigma_z\rangle=0$, i.e., the ferromagnetic order is confined to the
$XY$ plane on the Bloch sphere, thus confirming the expectation of a
$U(1)$ phase. The spontaneous magnetization thus simply fullows from
the length of the spin vector on the Bloch sphere. Investigating the
decay of the ferromagnetic order, we find that the $U(1)$ phase breaks
down at $\lambda_{c1} = 1/2$, undergoing a continuous transition, see
Fig.~\ref{fig:phase1}. Interestingly, incontrast to the equilbrium
case \cite{Mikeska2004,Fukumoto1996}, we do not find that the XY phase
is immediately connected to the antiferromagnet, but instead the
transition is to a disordered phase. From the variational analysis, it
appears that this novel intermediate phase is a paramagnet, however,
ordered phases involving a nonlocal order parameter cannot be ruled
out at this point.

\begin{figure}[b]
  \includegraphics[width=\linewidth]{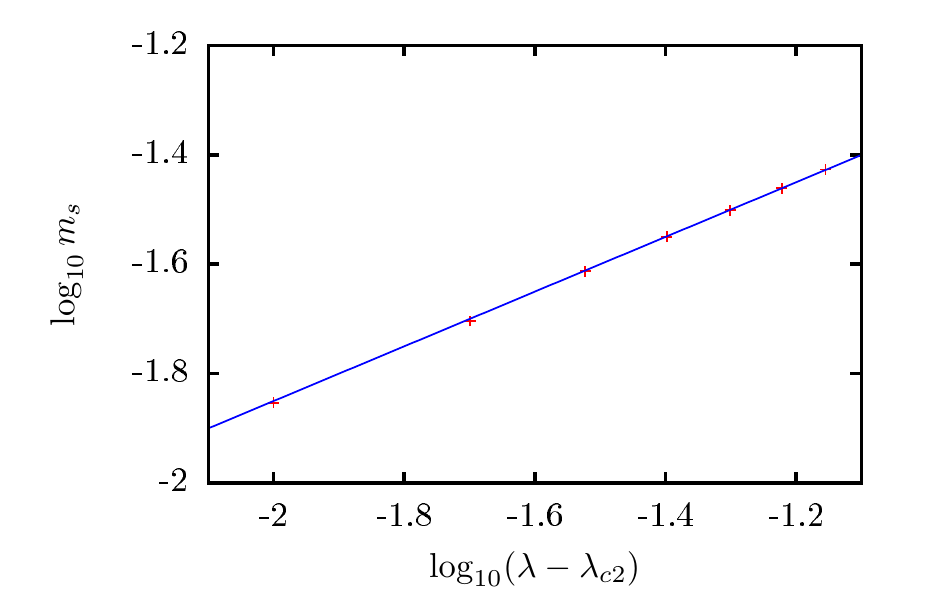}
  \caption{Critical behavior of the dissipative phase transition close $\lambda_{c2}=3/2$
    between the Ising antiferromagnet and the disordered phase. The fit represents a critical
    scaling of the order parameter indicating staggered magnetization according to $m_s \sim
    (\lambda-\lambda_{c2})^{1/2}$.}
\label{fig:phase2}
\end{figure}

\begin{figure}[t]
  \includegraphics[width=.7\linewidth]{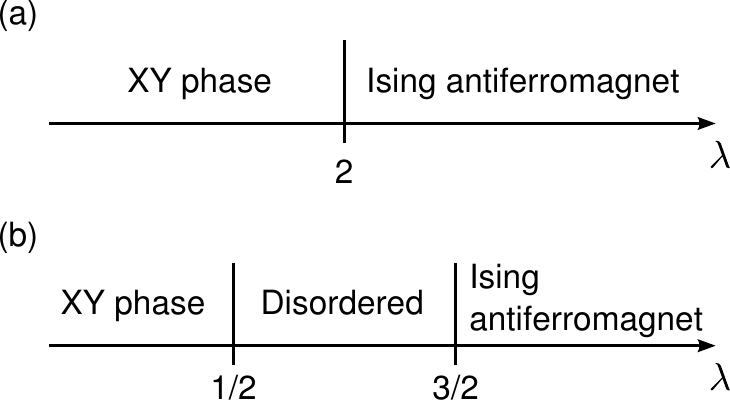}
  \caption{Phase diagram for the Heisenberg model. (a) Ground state
    phase diagram exhibiting a single transition from an XY phase to
    an Ising antiferromagnet. (b) Phase diagram of the dissipative
    model showing the presence of an additional disordered phase
    between the XY ferromagnet and the Ising antiferromagnet.}
  \label{fig:phase}
\end{figure}

I will now turn to the variational analysis for large values of
$\lambda$. In the limit $\lambda \to \infty$, the system has two dark
states corresponding to the antiferromagnetic ordering of the
spins. To incorporate antiferromagnetic ordering into the variational
approach is it necessary to consider variational states of the form
\begin{equation}
  \rho = \rho_A \otimes \rho_B \otimes \rho_A \otimes \rho_B \otimes \cdots.
\end{equation}
The amount of antiferromagnetic ordering is captured in the
staggered magnetization $m_s$, which is reduced for finite values of
$\lambda$. Again, we find a continuous phase transition at
$\lambda_{c2} = 3/2$ to the same disordered phase, see
Fig.~\ref{fig:phase2}. For both continuous transitions, the effective
critical theory appears to be a $\phi^4$ theory, resulting in the
Landau critical exponent of $\beta = 1/2$. The full phase diagram of
the dissipative Heisenberg model, including a comparison to the
equilbirium phase diagram, is shown in Fig.~\ref{fig:phase}.

The appearance of the intermediate disordered phase can be understood
from the fact that the situation in open systems is quite different to
finding the ground state of an equilbrium model. When constructing a
variational product state solution to a ground state problem, one can
never lower the variational energy by going from pure to mixed
states. Hence, all variational product states are pure and can be
characterized by unit vectors on the Bloch sphere. However, in open
systems it is well possible that going from pure states to mixed
states leads to a better approximation of the steady state. In this
case, the length of the Bloch vector is reduced until it vanishes
completely in a disordered phase. This is especially likely if the
system is far away from an exact dark state solution, which is the
case around $\lambda \approx 1$ for the dissipative Heisenberg
model. However, in contrast to the situation in \cite{Lammers2016},
the fully unpolarized state is never an exact steady state of the
quantum master equation, leaving the possibility for more exotic
ordering mechanisms to be realized within the dissipative Heisenberg
model.

\section{Summary}

I have presented a generic way to realize a large class of dissipative
quantum many-bodel spin models, including purely dissipative model
without any coherent dynamics. By performing a variational analysis of
a purely dissipative Heisenberg model, I have shown that the steady
state phase diagram exhibits an intermediate disordered phase that is
not present in the ground state phase diagram. These findings
underscore that the behavior of dissipative quantum many-body systems
is potentially richer than in equilibrium, and that the recently
introduced variational principle is very well suited to treat such
dissipative quantum many-body problems. In future developments, it
will be possible to adapt the variational principle to nonlocal
variational parameters in order to analyze whether the intermediate
phase is characterized by an exotic hidden order parameter.

\begin{acknowledgments}

  This work was funded by the Volkswagen Foundation and by the DFG
  within SFB 1227 (DQ-mat).

\end{acknowledgments}



\end{document}